\documentclass[10pt]{article}
\usepackage{graphicx}

\def\Title#1{\begin{center} {\Large #1 } \end{center}}
\def\Author#1{\begin{center}{ \sc #1} \end{center}}
\def\Address#1{\begin{center}{ \it #1} \end{center}}

\newcommand\pubblock{\rightline{\begin{tabular}{l} Proceedings of the Fifth Annual LHCP\\ \pubnumber\\
         \pubdate  \end{tabular}}}

\newenvironment{Abstract}{\begin{quotation} \begin{center} 
             \large ABSTRACT \end{center}\bigskip 
      \begin{center}\begin{large}}{\end{large}\end{center} \end{quotation}}

\newenvironment{Presented}{\begin{quotation} \begin{center} 
             PRESENTED AT\end{center}\bigskip 
      \begin{center}\begin{large}}{\end{large}\end{center} \end{quotation}}

\def\beq{\begin{equation}}
\def\eeq#1{\label{#1}\end{equation}}
\def\eeqn{\end{equation}}

\def\beqa{\begin{eqnarray}}
\def\eeqa#1{\label{#1}\end{eqnarray}}
\def\eeqan{\end{eqnarray}}

\let\bar=\overbar

\def\Dslash{\not{\hbox{\kern-4pt $D$}}}
\def\dslash{\not{\hbox{\kern-2pt $\del$}}}

\def\msb{{\bar{\ssstyle M \kern -1pt S}}}

\usepackage{color}
\usepackage{lineno}

\newcommand\pubnumber{ }

\newcommand\pubdate{\today}

\def\affiliation{
On behalf of the ALICE Collaboration, \\
Department of Physics - Torino University \\
and INFN Sezione di Torino, Torino, Italy}

\begin{document}
\large
\begin{titlepage}
\pubblock

\vfill
\Title{  PARTICLE IDENTIFICATION PERFORMANCE AT ALICE  }
\vfill

\Author{ ELENA BOTTA  }
\Address{\affiliation}
\vfill
\begin{Abstract}
ALICE (A Large Ion Collider Experiment) is the LHC experiment dedicated to the investigation of the nature and the properties of the Quark--Gluon Plasma (QGP) using heavy-ion collisions. Among its characteristics, excellent particle identification capabilities stand out, which are a basic requirement to address QGP physics. The ALICE particle identification performance will be reviewed in this paper, focusing in particular on the LHC Run 2 period.
\end{Abstract}
\vfill

\begin{Presented}
The Fifth Annual Conference\\
 on Large Hadron Collider Physics \\
Shanghai Jiao Tong University, Shanghai, China\\ 
May 15-20, 2017
\end{Presented}
\vfill
\end{titlepage}
\def\thefootnote{\fnsymbol{footnote}}
\setcounter{footnote}{0}
%

\normalsize 


\section{Introduction}
Heavy-ion collisions at the LHC energies produce high energy density and temperature conditions sufficient to produce a state of deconfined quarks and gluons, called Quark--Gluon Plasma (QGP) \cite{ref:qgp}. Transition to the ordinary colourless hadronic matter occurs due to its expansion and cooling. 

ALICE is a general-purpose detector designed to address the physics of strongly interacting matter and the QGP by means of a comprehensive study of hadrons, electrons, muons, and photons produced in the collision of heavy nuclei (Pb--Pb), up to the highest multiplicities achieved at the LHC energies. Particle identification (PID) capabilities are thus essential for a comprehensive study of the final hadronic state and the achievement of excellent performance underlay the detector design. 
In Section 2 of this paper the ALICE detector will be briefly described. The PID techniques used by the experiment and the performance obtained in the LHC Run 2 will be presented in Section 3. Conclusions will be drawn in Section 4.

\vspace{-2mm}
\section{The ALICE detector}
\vspace{-1mm}
\begin{figure}[htb]
\centering
\includegraphics[height=2.1in]{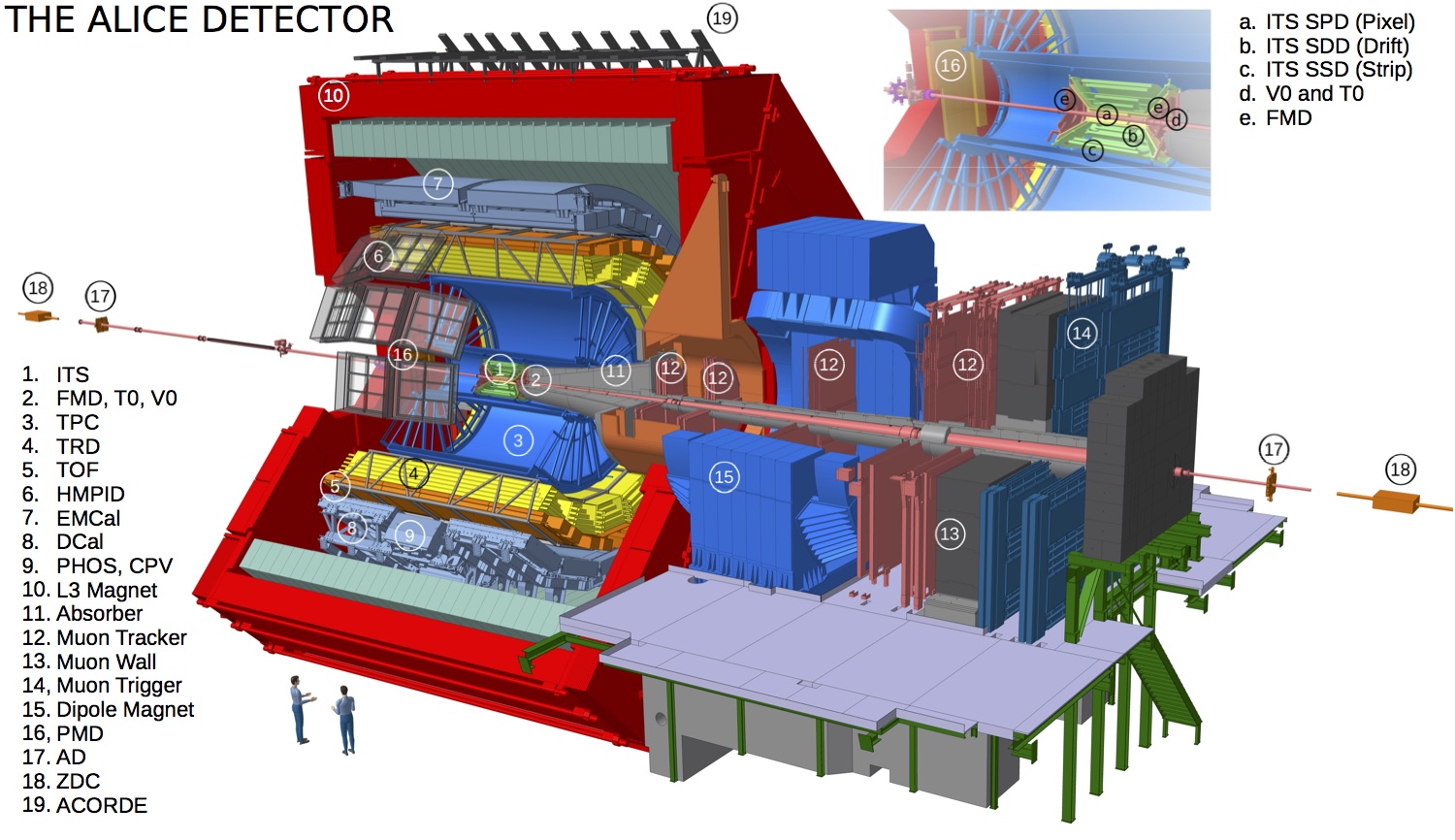}
\hspace{2mm}
\includegraphics[height=1.6in]{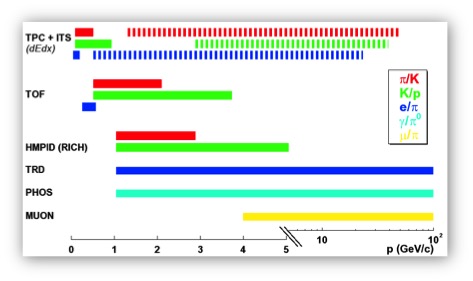}
\caption{Left panel: schematic view of the ALICE detector. Right panel: pictorial overview of the momentum intervals in which different ALICE subsystems provide PID for hadrons, electrons muons, and photons.}
\label{fig:fig1}
\end{figure}

Figure \ref{fig:fig1} shows a scheme of the ALICE apparatus in the LHC Run 2. 
The ALICE detector ensemble can be divided into central barrel and single arm detectors, based on their full or partial azimuth coverage, forward detectors and the MUON spectrometer. Moving radially outward from the beam pipe the central-barrel detectors – Inner Tracking System (ITS), Time Projection Chamber (TPC), Transition Radiation Detector (TRD), Time-Of-Flight (TOF) are located; the single arm Photon Spectrometer (PHOS), Electromagnetic Calorimeter (EMCal), Di-Jet calorimeter (DCal) and High Momentum Particle Identification Detector (HMPID) surround them. They are all embedded in the L3 solenoid magnet which has B=0.5 T.
At forward rapidities a muon spectrometer (MUON) is placed.
Some more detectors complete the ALICE setup, but they won't be described since they will not be  considered in this paper. For more details about the ALICE detectors see \cite{ref:jinst}.

The ITS and the TPC are the main charged-particle tracking detectors \cite{ref:perf}. The ITS is composed of six tracking layers, two Silicon Pixel Detectors (SPD), two Silicon Drift Detectors (SDD), and two Silicon Strip Detectors (SSD). The TPC has a 90 m$^{3}$ drift volume, which was  filled with 
Ar-CO$_{2}$ in 2015--2016 data takings, divided into two parts by the central cathode. The end plates are equipped with multiwire proportional chambers (MWPC). The ITS and TPC are used for the reconstruction of the primary and secondary vertices and also permit the reconstruction of low-momentum tracks. The ITS is capable to perform standalone reconstruction to recuperate the tracks lost in the global tracking due to the spatial acceptance, the intrinsic momentum cutoff of the outer detectors, and to particle decays. The TPC has a tracking efficiency of $\sim$ 80$\%$ in $|\eta |< 0.8$ region up to $p_{\mathrm{T}}$ = 10 GeV/{\it c}, with a momentum resolution $\sigma_{p_{\mathrm{T}}}/p_{\mathrm{T}} \sim 5 \%$ which almost halves when combined with ITS. In addition to tracking, SDD, SSD and TPC provide charged-particle identification via measurement of the specific ionization energy loss d$E$/d$x$.
The TRD consists of six layers of Xe–CO$_{2}$-filled MWPCs preceded by a drift region of 3 cm, with a fiber/foam radiator in front of each chamber. It is used for charged-particle tracking and for electron identification via transition radiation and d$E$/d$x$. The TOF detector is based on Multigap Resistive Plate Chamber (MRPC) technology; it is located at a radius of 3.7 m from the interaction point and provides particle identification at intermediate momenta.
Finally, the cylindrical volume outside TOF is shared by three electromagnetic calorimeters with thicknesses of $\sim$20 {\it X}$_{0}$ (radiation lengths) and $\sim$1 $\lambda_{int}$ (nuclear interaction length): the high-resolution PHOS, the large-acceptance EMCal and DCal, which form a two-arm electromagnetic calorimeter for di-jet detection, along with the ring-imaging Cherenkov detector HMPID, which has a liquid C$_{6}$F$_{14}$ radiator and a CsI photo-cathode for charged-hadron identification at intermediate momenta.

For completeness, it must be noted that two subsystems, DCal and AD, the ALICE Diffractive Detector designed to study pp diffractive processes at $|\eta|$ up to 12.1, have been put in operation and also the azimuthal coverage of the TRD has been completed for the LHC Run 2 in 2015.

\section{PID in ALICE: techniques and performance}
Many of the ALICE detector subsystems provide particle identification information, using all the PID techniques known nowadays, in their state of art implementation. A synopsis is given in Table \ref{tab:table1}.

\begin{table}[h]
\begin{center}
\begin{tabular}{ll|ll}
\multicolumn{2}{c|}{\bf{Central barrel subsystem}} & \multicolumn{2}{c}{\bf{Single arm subsystem}} \\
\hline
\bf{Technique} & \bf{detector} & \bf{Technique} & \bf{detector}  \\
\hline 
 Specific energy loss & ITS, TPC &  Calorimetry & PHOS   \\
& TRD & & EMCal, DCal \\
\hline
Transition radiation & TRD &  Muon spectrometry & MUON arm \\
\hline
Time of Flight & TOF & Cherenkov radiation & HMPID  \\
\hline
\end{tabular}
\caption{PID techniques provided by the ALICE barrel and single arm detector subsystems.}
\label{tab:table1}
\end{center}
\end{table}
\begin{table}[h]
\begin{center}
\vspace{-5mm}
\begin{tabular}{ll}
 \multicolumn{2}{c}{\bf{Decay topology}} \\
\hline
\bf{particle/method} & \bf{detector} \\
\hline 
  K$^{\pm}$ via kinks & TPC \\
\hline
$K^{0}_{S}$, $\Lambda$, $\Xi ^{-}$, D & ITS, TPC \\
\hline
$\pi^{0}$, $\eta$ via photon conversion (PCM) & ITS, TPC, TOF \\
\hline
\end{tabular}
\caption{Decay topology PID methods implemented by the ALICE barrel detector subsystems.}
\label{tab:table2}
\end{center}
\vspace{-1mm}
\end{table}
\vspace{-1mm}
The central barrel detectors provide PID using the specific ionization energy of a charged particle traversing a medium, the transition radiation emitted by charged particles when crossing the boundary between two materials for $\gamma >$1000, and the time of flight a charged particle takes to reach a detector sensitive volume from the interaction point.

The d$E$/d$x$ measurements are provided by the four outermost layers of the ITS detector, SDD and SSD, featuring an  analog readout. A truncated mean is applied to the measurements: an average of the lowest two is taken if all the four signals are there, or a weighted average is taken if only three are available. The ITS PID is performed on a track-by-track basis in the low {\it p}$_{\mathrm{T}}$ region with a resolution of 10--12 $\%$, up to $\sim$ 1 GeV/{\it c}, and pions reconstructed in standalone mode can be identified down to  {\it p}$_{\mathrm{T}}$=100 MeV/{\it c}. 

\begin{figure}[htb]
\vspace{-4mm}
\centering
\includegraphics[height=2.in]{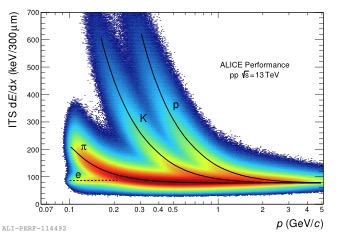}
\hspace{2mm}
\includegraphics[height=2.0in]{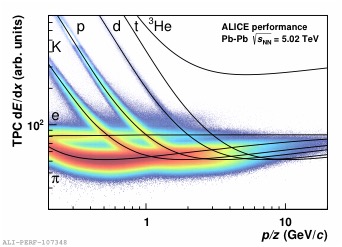}
\caption{Left panel: specific energy loss measured by the ITS as a function of $p_{T}$. Right panel: specific energy loss measured by the TPC as a function of rigidity for positive tracks. The curves represent parameterizations of the expected mean energy loss.}
\label{fig:fig2}
\end{figure}

Specific energy loss measurements are provided by the ALICE TPC detector as well. Also in this case, a truncated mean is applied over the maximum number of 159 cluster information. The performance is excellent, with a resolution of $\sim 5.2 \%$ for pp collisions and $\sim 6.5 \%$ for $0-5 \%$ most central Pb--Pb collisions. Charged hadron PID is performed on a track-by-track basis up to 1--2 GeV/{\it c}. The wide dynamic range (up to 26 MIP) allows to identify light nuclei, like deuterons, tritons and $^{3}$He. Moreover, in the region of the relativistic rise a statistical approach is utilized, allowing the TPC to identify charged hadrons up to $p_{T}$ of a few tens of GeV/{\it c}.
The specific energy loss distributions measured in LHC Run 2 are shown in Figure \ref{fig:fig2} for the ITS on the left in pp collisions as a function of $p_{T}$ and for the TPC on the right in Pb--Pb collisions as a function of the rigidity $p/z$. 

\begin{figure}[h]
\vspace{-1mm}
\centering
\includegraphics[height=1.7in]{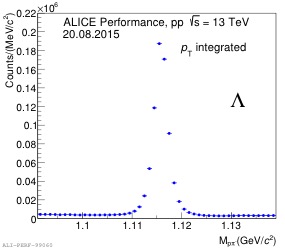}
\hspace{2mm}
\includegraphics[height=1.7in]{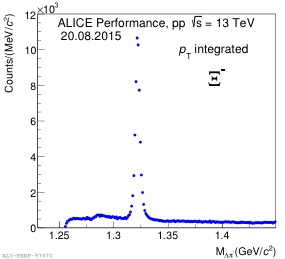}
\hspace{2mm}
\includegraphics[height=1.7in]{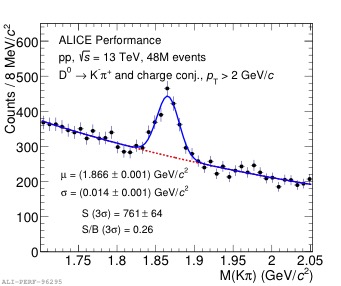}
\caption{Left panel: $p \pi^{-}$ invariant mass in the $\Lambda$ region; central panel: $\Lambda \pi^{-}$ invariant mass in the $\Xi^{-}$ region; right panel: $K^{-} \pi^{+}$ invariant mass in the $D_{0}$ region. All plots refer to LHC Run 2 data taking.}
\label{fig:fig3}
\end{figure}

PID can also be performed in ALICE by exploiting peculiar decay topologies of long-lived charged and neutral particles, as indicated in Table \ref{tab:table2}. 
Charged kaons can be identified in the TPC by a distinct kink in the track owing to the decay into a muon and a neutrino or into a charged and a neutral pion pair. $K^{0}_{S}$ and $\Lambda$ can be identified via the so-called "V0 topology" of the charged daughter particles tracks reconstructed in the ITS and TPC. $\Xi^{-}\rightarrow \Lambda \pi^{-}$ and $D^{0}\rightarrow K^{-} \pi^{+}$ can be identified by their decay topology as well and also the $^{3}_{\Lambda}$H hypernucleus can be detected by its $^{3}$He+$\pi^{-}$ decay \cite{ref:3LH}. Figure \ref{fig:fig3} shows examples of  invariant mass distributions obtained from the 2015 data taking. 

Electron identification in ALICE is provided by the TRD in the momentum region {\it p} $>$ 1 GeV/{\it c}, with a pion rejection factor of about 100 \cite{ref:perf}. The PID relies on a 1-dimentional likelihood approach, which makes it possible to distinguish between pions and electrons due to the different shapes of the signals induced in the detector due to the transition radiation. A 2-dimensional likelihood method is also used: it is based on the temporal distribution of the signal and provides further improvement of the pion rejection.  

\begin{figure}[h]
\centering
\includegraphics[height=2.0in]{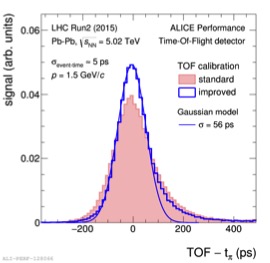}
\hspace{2mm}
\includegraphics[height=2.0in]{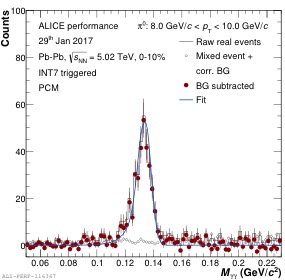}
\caption{Left panel: TOF detector time resolution given by the difference between the measured time of flight and the expected one calculated from the track length and momentum assuming that the particle is a pion. 
Right panel: $\pi^{0}$ invariant mass distribution obtained with the photon conversion method (described in the following) in Pb--Pb collisions at $\sqrt{s_{NN}}$=5.02 TeV .
}
\label{fig:fig4}
\end{figure}

Charged hadrons in the intermediate momentum range (i.e. up to a few GeV/{\it c}) are identified in ALICE by the TOF detector \cite{ref:perf}. In this case, the mass (and thus the identity) of a particle is obtained by combining the measurement of its time of flight (from TOF) and its momentum (from ITS and TPC). The reference time of the event is given on an event-by-event basis by the TOF, by the ALICE T0 detector or by the combination of their information depending on the collision system \cite{ref:tof}.
The left panel of Figure \ref{fig:fig4} shows the TOF resolution for the identification of pions (the most abundant particle species) in terms of the difference between the measured time of flight and the expected one calculated from the track length and momentum assuming that the particle is a pion. As one can see, the detector performance is outstanding, with a resolution $\sigma\sim$ 56 ps obtained thanks to a refined time slewing calibration procedure.
The $\pi/K$ and $K$/p separation is obtained at a $3 \sigma$ level up to 2.5 GeV/{\it c} and 4 GeV/{\it c}, respectively, on a track-by-track basis. At higher transverse momenta where the TOF does not permit track-by-track identification, a fit of multiple Gaussian peaks is used to determine the particle yields up to $p_{\mathrm{T}}$ of $\sim$5 GeV/{\it c}.

The TPC and TOF PID information can be used in conjunction, e.g. by combining the n-$\sigma$ identification requirements: this method is very effective for electrons at transverse momenta as high as 8 GeV/{\it c} where single detectors do not allow a clean separation.

Also neutral mesons like $\pi^{0}$ and $\eta$ can be reconstructed through their $\gamma \gamma$ decay by detecting both the $\gamma \rightarrow e^{-}e^{+}$ conversion reactions in materials: this method is called Photon Conversion Method (PCM) \cite{ref:perf}. 
The converted photon and its conversion point can be reliably measured by reconstructing the electron and positron with the ITS, TPC and TOF for conversions within 180 cm from the beam axis, by reconstructing the particular V0 topology. The main sources for conversions are the beam pipe, the ITS layers, the TPC vessel and drift gas. The ITS services and the ITS and TPC support structures give additional contributions. Being the photon conversion probability very sensitive to the amount, geometry, and chemical composition of the traversed material, it is vital to have accurate knowledge of the material budget to apply the PCM. The right panel of Figure \ref{fig:fig4} shows the  $\pi^{0} \rightarrow \gamma \gamma$ invariant mass spectrum obtained in Run 2 with Pb--Pb collisions at $\sqrt{s_{NN}}$ = 5.02 TeV.
 
Hadron identification at intermediate momenta (up to 3--5 GeV/{\it c} depending on the species) is performed using the HMPID detector \cite{ref:perf}. This is a single-arm proximity focusing RICH, which determines the $\beta$ of a particle from the measurement of the Cherenkov angle. This information is then combined with the momentum measured by the TPC and ITS to assign an identity to the particle.

The different PID methods provided by ITS standalone, TPC-TOF, TOF, HMPID and kink identification allow to perform identification in different transverse momentum regions which partially overlap. Since  in these regions the spectra from the different PID techniques are consistent within uncertainties, 
a combined analysis procedure has also been developed \cite{ref:comb} in which 
they are sequentially averaged starting from the two analyses whose results are the most closely correlated (namely TPC-TOF and TOF) and then adding the others one-by-one according to their degree of correlation with the previous ones. Identified $p_{T}$ spectra of $\pi$, K and p were obtained from the data of the LHC Run 1; for Run 2 data a similar analysis is ongoing.

A Bayesian approach method has also been developed to combine PID signals from ITS, TPC and 
TOF, which makes effective use of the full PID capabilities of ALICE \cite{ref:bayes}. It features a recursive procedure to determine the priors probabilities to measure each particle species. The method has been validated by comparing its results to those of a variety of analyses performed with different PID techniques: its suitability has been fully tested and assessed. 

The three ALICE electromagnetic calorimeters, PHOS, EMCal and DCal measure $\gamma$ up to 100 and 250 GeV respectively. The EMCal and DCal are also used in ALICE to help to reject hadrons when identifying electrons, thanks to the {\it E/p} distribution characteristically peaked at 1 only for electrons due to their small mass \cite{ref:perf}. 

\begin{figure}[h]
\centering
\includegraphics[height=1.7in]{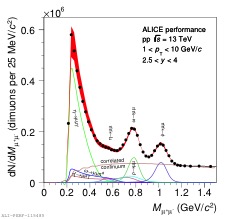}
\hspace{2mm}
\includegraphics[height=1.7in]{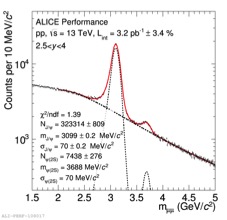}
\hspace{2mm}
\includegraphics[height=1.7in]{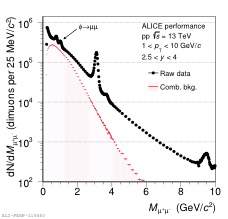}
\hspace{2mm}
\caption{
Dimuon invariant mass spectra obtained by the ALICE MUON detector in the LHC Run 2. Left panel: 0--1.5 GeV/$c^{2}$ interval; central panel: 1.5--5 GeV/$c^{2}$ interval; right panel: 0--10 GeV/$c^{2}$ interval.}

\label{fig:fig6}
\end{figure}

On one side of the experiment, a MUON spectrometer reconstructs and identifies muons in the momentum range {\it p} $>$ 4 GeV/{\it c} and in the $4 < \eta <2.5$ range \cite{ref:perf}. Hadron rejection is performed by matching tracks between the tracking chambers and the triggering chambers. Moreover, geometrical and topological cuts are applied in order to reduce contamination from fake tracks, and Monte Carlo simulations are used to estimate the muon contributions from hadron decays. $\eta, \omega,\rho,\phi, J/\psi, \psi(2S)$, $Y$ mesons can be identified through their decay into $\mu^{+} \mu^{-}$ pairs.  
Figure \ref{fig:fig6} shows the invariant mass distribution of dimuons (0 $<p_{T} <$ 10 GeV/{\it c}) reconstructed and identified by the MUON spectrometer in the LHC Run 2. 

\vspace{-6mm}
\section{Conclusions}
PID capabilities are a unique feature of the ALICE experiment.
ALICE features many powerful detectors exploiting all known PID techniques for charged and neutral particles, all of them showing a high-level technology implementation. 
A wide momentum range is covered by different PID techniques allowing particle separation up to the highest multiplicities obtained at the LHC energy scale. 
The availability of PID signals from many detectors has led to the development of tools for multi-detector approaches. 
More than 50$\%$ of the LHC Run 1 ALICE publications involve PID techniques: for Run 2 data all the PID tools have demonstrated to be well in shape and will allow many new interesting physics results to be published soon.





\end{document}